# Abstract art grandmasters score like class D amateurs

Mikhail Simkin

Hawley-Dolan and Winner had asked the art students to compare paintings by abstract artists with paintings made by a child or by an animal. In 67% of the cases, art students said that the painting by a renowned artist is better. I compare this with the winning probability of the chessplayers of different ratings. I conclude that the great artists score on the level of class D amateurs.

Hawley-Dolan and Winner [1] had shown to 32 art students 30 pairs of paintings. In each pair, one painting was by a renowned abstract expressionist and another by a child or by an animal (monkey, gorilla, chimpanzee, or elephant). They asked the subjects which painting is better. In the case when the paintings were unlabeled, the art students in 67% of the cases said that the painting by a renowned artist is better. Hawley-Dolan and Winner concluded that their findings "challenge the common claim that abstract expressionist art is indistinguishable from (and no better than) art made by children."

But how much is it better? Four years ago, I reported [2] the results of my "True art, or fake?" quiz [3]. It consists of a dozen pictures. Some of them are masterpieces of abstract art, created by famous artists. The rest are fakes, produced by myself. The takers are to tell which is which. The average score received by fifty-six thousands quiz-takers is 7.91 out of 12 or 65.9% correct. A figure close to the one reported by Hawley-Dolan and Winner. What does it say about the difference in quality? In [2] I answered the question using a comparison with one classic study [4].

A hundred years ago psychologist F.M. Urban performed an experiment on just perceptible differences [4]. He asked the subjects to compare a hundred-gram weight with a set of different weights. When the weights were close, the judgment was poor. However, statistically, the lighter weight appeared to be heavier in less than half of the cases. For example, a 100-gram weight appeared to be heavier than 96 and 108-gram weights in 72% and 9% of trials respectively.

An abstractionist is judged better than a child/animal in 67% of the trials, while a 100-gram weight is judged heavier than 96-gram weight in 72% of the trials. This means that there is less perceptible difference between an abstractionist and child/animal than between 100 and 96 gram. Thus, if we assign to the artistic heavyweights used in [1] the weight of 100 artistic grams the weight of child/animal paintings is more than 96 artistic grams. The difference in their weights is 4%.

In the sport of weightlifting participants are divided into eight categories according to their weight [5]. The boundaries of bordering categories are about 12% apart. For example, people weighting between 94 and 105 kg belong to the same weight category. I must conclude that children and monkeys either belong to the same weight category with abstract expressionist heavyweights or are just one category below.

More simple and instructive is the comparison with chess. In 1960$^{th}$ Arpad Elo developed a rating system for chessplayers which is now in common use [6]. Elo rating is a number based on player's performance, which can be used to compare players' relative strength. When the difference in rating between the players is 200 points, the probability that a higher rated player will win the game is 77%. When the difference is 400 or 600 the corresponding probabilities are 93% and 98% respectively. The players are divided into nine categories (see Table 1), from world championship contenders to novices, with the difference between bordering categories of 200 points.

Since abstract expressionist wins over a monkey only in 67% of cases, the difference in their artistic Elo ratings is less than 200. This means that they either belong to the same category with apes or are just one category above. If we class a gorilla as a novice, abstract art grandmasters are at best class D amateurs.

**Table 1.** Elo rating of chessplayers. When the difference in rating between the players is 200 points, the probability that a higher rated player will win the game is 77%.

| Elo rating | Player category |
| --- | --- |
| 2600 and up | World championship contenders |
| 2400 - 2600 | Grandmasters |
| 2200 - 2400 | National masters |
| 2000 - 2200 | Candidate masters |
| 1800 - 2000 | Amateurs – Class A |
| 1600 - 1800 | Amateurs – Class B |
| 1400 - 1600 | Amateurs – Class C |
| 1200 - 1400 | Amateurs – Class D |
| Below 1200 | Novices |